\documentclass[
superscriptaddress,
twocolumn,
 amsmath,amssymb,
 aps,
prb,
]{revtex4-1}
\usepackage{graphicx,color}
\usepackage{dcolumn}
\usepackage{bm}
\usepackage{physics}
\begin{document}
\preprint{cond-mat}
\title{Algebraic charge dynamics of quantum spin-liquid $\beta'$-EtMe$_{3}$Sb[Pd(dmit)$_{2}$]$_{2}$}
\author{Shigeki\ Fujiyama}
\email[E-mail: ]{fujiyama@riken.jp}
\author{Reizo\ Kato}
\affiliation{RIKEN, Condensed Molecular Materials Laboratory, Wako 351-0198, Japan}
\date{12 December 2017}
\begin{abstract}
Nuclear spin-lattice ($1/T_{1}$) and spin-spin ($1/T_{2}$) relaxation rates of the cation sites of a candidate quantum spin-liquid $\beta'$-EtMe$_{3}$Sb[Pd(dmit)$_{2}$]$_{2}$ and its deuterated sample are presented. The enhanced $1/T_{1}$ of $^{1}$H and $^{2}$D is thoroughly analyzed considering the rotations of methyl and ethyl groups of the cation with activation energies of $2\times 10^{2}$ K and $1.2 \times 10^{3}$ K respectively. Contrasting charge dynamics with an algebraic temperature dependence is found at the Sb site in the EtMe$_{3}$Sb cation. The charge fluctuation remains active down to the lowest temperature, as has been observed in the ac response of dielectric constants. 
\end{abstract} 
\maketitle

\section{Introduction}
Quantum spin-liquids (QSLs) have attracted considerable attention in condensed matter physics.~\cite{Balents2010,Balents2002} Geometrical frustration of the antiferromagnetic network in triangular or Kagome lattices is known to prevent classical N\'eel-type magnetic ordering with two sublattices and stabilize novel magnetic states. Among the QSLs, the realization of a gapless QSL whose ground state is a direct product of spin singlets, as first proposed by Anderson,~\cite{Anderson1973} has been pursued for a long time, however, its materialization remains a challenging issue.

Several candidate organic salts have been extensively studied.~\cite{Kanoda2011} Of these, $\beta'$-EtMe$_{3}$Sb[Pd(dmit)$_{2}$]$_{2}$ (dmit = 1,3-dithiole-2-thione-4,5-dithiolate) and $\kappa$-(BEDT-TTF)$_{2}$Cu$_{2}$(CN)$_{3}$ are examined thoroughly by both magnetic and electronic measurements.~\cite{Tamura2002,Shimizu2003,Kurosaki2005,Shimizu2016} No evidence of classical antiferromagnetic order down to $T\sim 1/1000 J$ ($J$ is the antiferromagnetic coupling constant) has been reported so far. The estimated Wilson ratio ($R_{W}\equiv\chi/\gamma \sim 1.1 $ [emu J$^{-1}$K$^{2}$]), which is a measure of the enhancement of the mass of quasiparticles, suggests reasonably large surfaces for the spin (spinon) excitation.~\cite{MYamashita2010,SYamashita2008,SYamashita2011} These experimental results are consistent with the gapless long-range resonating valence bond (RVB) scenario.

Apart from theoretical concept of the novel state, the actual material raises fundamental problems related to macroscopic quantum phenomena. One is related to the stability or robustness of the QSL against the inhomogeneity of the electronic state inherent in real materials. Although there is little opportunity for an impurity to replace the atoms in molecules in the crystallization of the organic salts, we can exemplify several sources for perturbing electronic states. For example, counter cations [in the case of Pd(dmit)$_{2}$ salts] and anions [BEDT-TTF salts] that are electronically closed shell have freedom of quenching disorders, which can cause local electronic instability. The wide distribution of the electronic spin $S=1/2$ on [Pd(dmit)$_{2}$]$_{2}$ or (BEDT-TTF)$_{2}$ dimers also causes inhomogeneity. The tight-binding approach to compose a geometrical network of $S=1/2$ spins cannot take the intradimer degree of freedom into account.

It was previously found that the ac responses of the dielectric constants of the molecular-based QSLs $\beta'$-EtMe$_{3}$Sb[Pd(dmit)$_{2}$]$_{2}$ and $\kappa$-(BEDT-TTF)$_{2}$Cu$_{2}$(CN)$_{3}$ show similar anomalies.~\cite{AbdelJawad2010,AbdelJawad2013} The dielectric constants $\epsilon'$ of both materials show enhancement between 10 K and 50 K at low frequencies below 1 MHz. The resemblance of the dielectric response for very different counter ions composed of different elements (EtMe$_{3}$Sb and Cu$_{2}$(CN)$_{3}$, both are electronically closed) indicates that the low-energy dielectric fluctuations originate from dimerized molecules that hold $S=1/2$ spins. This is expected to disturb the geometrical network of the antiferromagnetic interaction, by which a massively degenerated ground state could be lifted. Recent theoretical works recognize the role of electric dipoles or charge fluctuations in stabilizing QSLs,~\cite{Hotta2010,Naka2010,Naka2015} and electronic interaction between magnetic (BEDT-TTF)$_{2}$ and anion layers is proposed as an origin for the observed ferroelectricity.~\cite{Dressel2016,Pinteric2016} However, the macroscopic dielectric constants cannot be compared with theoretical results, which limits quantitative understanding of the charge dynamics of QSL materials. In addition, it is known that the terminal ethylene groups in the BEDT-TTF molecule have degree of freedom in conformation, which can significantly affect the dielectric constants. Microscopic measurements of charge fluctuations have been awaited to unveil the charge dynamics of QSLs.

\begin{figure}[htb]
\includegraphics[width=8.5cm]{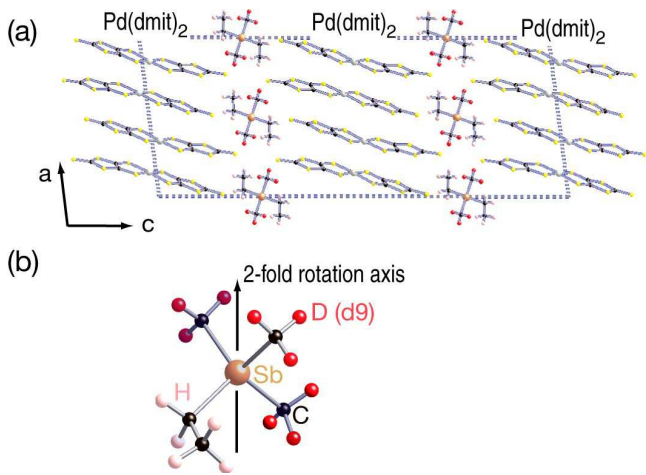}
\caption{(a) Crystal structure of $\beta'$-EtMe$_{3}$Sb[Pd(dmit)$_{2}$]$_{2}$. The [Pd(dmit)$_{2}$]$_{2}$ and cation layers are alternately stacked. (b) EtMe$_{3}$Sb cation. The Sb atom is located on the twofold rotation axis of the unit cell. In a d9-deuterated sample, protons of three methyl groups are labeled by deuterons.}
\label{fig:cation}
\end{figure}

In this article, we report the NMR results for $^{1}$H, $^{2}$D, and $^{121}$Sb in the cation of $\beta'$-EtMe$_{3}$Sb[Pd(dmit)$_{2}$]$_{2}$. The nuclear spin-lattice relaxation rates ($1/T_{1}$) and nuclear spin-spin relaxation rates ($1/T_{2}$) of the three elements have maxima caused by low-energy electronic fluctuations. We assign the temperature-dependent correlation times with activation energies of $\Delta = 2 \times 10^{2}$ and $\Delta =1.2 \times 10^{3}$ to the rotations of methyl (CH$_{3}$) and ethyl (C$_{2}$H$_{5}$) groups by comparing $1/T_{1}$ of $^{1}$H and $^{2}$D for pristine and d9-deuterated (C$_{2}$H$_{5}$(CD$_{3}$)$_{3}$Sb) crystals. At the Sb site, on the other hand, we found unknown glassy slowing down of the charge fluctuation with a weakly temperature-dependent correlation time, which remains active down to the lowest temperatures.

\section{Experimentals}
Single crystals of $\beta'$-\textit{X}[Pd(dmit)$_{2}$]$_{2}$ (\textit{X} = EtMe$_{3}$Sb and Me$_{4}$Sb cations) were prepared by air oxidation of \textit{X}$_{2}$[Pd(dmit)$_{2}$] in an acetone solution containing acetic acid.~\cite{Kato2012} We show in Fig.~\ref{fig:cation} the crystal structure of \textit{X} = EtMe$_{3}$Sb. There is a twofold rotation axis of the unit cell through Sb site. NMR measurements were performed using small randomly oriented single crystals. The resonating Larmor frequencies, $\omega_{0}$, used were 125.36 MHz for $^{1}$H, 46.5 MHz for $^{2}$D, and 79.2 MHz for $^{121}$Sb NMR. $1/T_{1}$ are obtained by fitting the magnetization, $M(t)$, by formulae, $M(t)/M_{\infty}= \mathrm{e}^{-t/T_{1}}$ for $^{1}$H, $M(t)/M_{\infty}= 1/4 ~\mathrm{e}^{-t/T_{1}}+3/4~\mathrm{e}^{-3t/T_{1}}$ for $^{2}$D, and  $M(t)/M_{\infty}= 9/35~ \mathrm{e}^{-t/T_{1}}+4/15~\mathrm{e}^{-6t/T_{1}}+10/21~ \mathrm{e}^{-15t/T_{1}}$ for $^{121}$Sb NMR, respectively.

\section{Results and Discussion}
\begin{figure}[hbt]
\includegraphics[width=8.5cm]{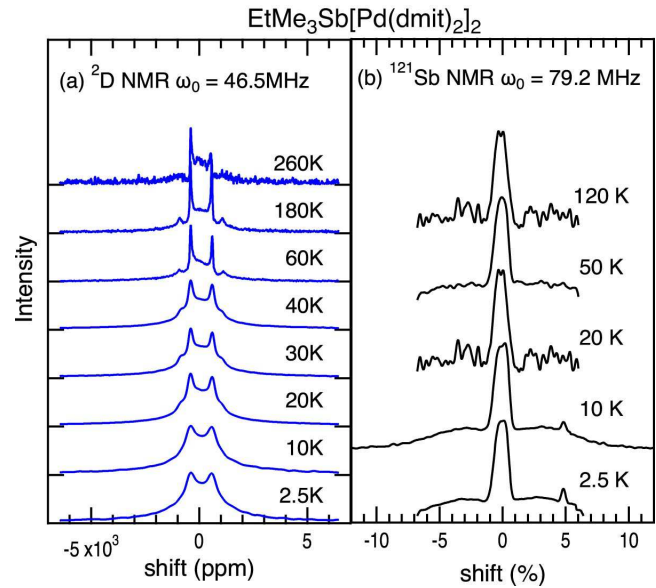}
\caption{NMR spectra of (a) $^{2}$D of d9-deuterated and (b) $^{121}$Sb of pristine samples of EtMe$_{3}$Sb[Pd(dmit)$_{2}$]$_{2}$. In Fig.~\ref{fig:Dspectra} (b), $^{63}$Cu signal from NMR coil is visible at $\sim 5$ \%.}
\label{fig:Dspectra}
\end{figure}

We show in Fig.~\ref{fig:Dspectra} the $^{2}$D and $^{121}$Sb NMR spectra of d9-deuterated EtMe$_{3}$Sb[Pd(dmit)$_{2}$]$_{2}$. The electric field gradient of $^{2}$D is found to be axially symmetric by Fig.~\ref{fig:Dspectra} (a). This is consistent with the crystal structure, and we consider that the axis of the local C$_{3}$ symmetry of the CD$_{3}$ methyl group is the quantized axis of the electric field gradient of $^{2}$D. Upon lowering the temperature, the spectra show significant broadening below 40 K indicating freezing of the rotation of the methyl group. Large nuclear quadrupole moment, $Q$, of $^{121}$Sb compared that of $^{2}$D with $|^{121}Q|/|^{2}Q|>10^{2}$ causes significant broadening of the spectra as shown in Fig.~\ref{fig:Dspectra} (b), and no notable change is found for whole temperature. At $\omega_{0}$, the dominating source to contribute the NMR signal is the central transition, which enabled us to obtain $1/T_{1}$ with small error.

\begin{figure}[htb]
\includegraphics*[width=8.5cm,trim=0 0 0 0]{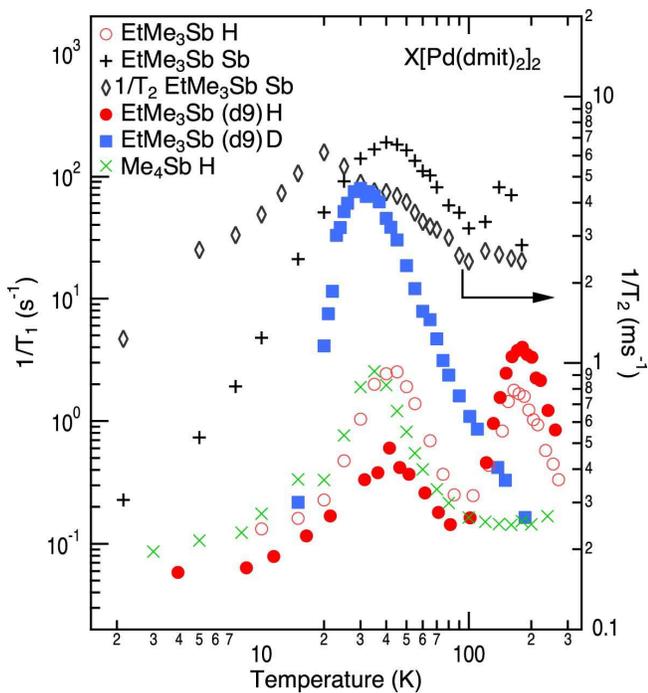}
\caption{Nuclear spin-lattice relaxation rates ($1/T_{1}$) of $^{1}$H and $^{121}$Sb for the pristine sample and those of $^{1}$H and $^{2}$D for the d9-deuterated sample. We plot $1/T_{1}$ for \textit{X} = Me$_{4}$Sb as reference data of a cation with tetrahedral symmetry. The black diamonds $\diamondsuit$ show the nuclear spin-spin relaxation rate ($1/T_{2}$) of Sb for the pristine sample measured at 79.2 MHz.}
\label{fig:iT1Tdep}
\end{figure}

We plot in Fig.~\ref{fig:iT1Tdep} the $1/T_{1}$ of $^{1}$H, $1/T_{1}$ and $1/T_{2}$ of $^{121}$Sb for the pristine EtMe$_{3}$Sb cation, and $1/T_{1}$ of $^{1}$H and $^{2}$D for the d9-deuterated cation. We also plot $1/T_{1}$ of $^{1}$H for \textit{X} = Me$_{4}$Sb, which undergoes antiferromagnetic ordering at $T_\mathrm{N}=11$ K, as reference data. Each data set strongly depends on the temperature, and we consider the motion in the cation as the dominant source enhancing nuclear relaxations.

In molecular solids, the rotations of methyl and ethyl groups cause magnetic and charge fluctuations of the local fields. In the case of \textit{X}[Pd(dmit)$_{2}$]$_{2}$, the rotations of both groups in cation \textit{X} are the sources, even though the cations are electronically closed. The fluctuation of local fields enhances $1/T_{1}$ as follows:~\cite{Dimitropoulos1990}
\begin{equation}
\frac{1}{T_{1}}=\expval{\Delta \omega^{2}} \qty[\frac{\tau_{c}}{1+\omega_{0}^{2}\tau_{c}^{2}}+\frac{4\tau_{c}}{1+4\omega_{0}^{2}\tau_{c}^{2}}],
\label{eq:BPP}
\end{equation}
where $\expval{\Delta \omega^{2}}$ is the variance of the fluctuation, which depends on the nuclear relaxation process, $\tau_{c}$ is the temperature dependent correlation time, and $\omega_{0}$ is the Larmor frequency, which is fixed for each NMR measurement. The $1/T_{1}$ shows a maximum at $\omega_{0} \tau_{c}=0.62$.

For $^{1}$H NMR, the relaxation is dominated by the H-H intradipolar interaction which is perturbed by rotations of methyl and ethyl groups, from which we obtain
\begin{equation}
\expval{\Delta \omega^{2}}=\frac{2}{5}\qty(\frac{\mu_{0}}{4\pi})\frac{\gamma^{4}_{N}\hbar^{2}}{r^{6}}I(I+1)
\end{equation}
where $\mu_{0}$ is the Bohr magneton, $\gamma_{N}$ is the gyromagnetic ratio of the nuclei, $r$ is the nearest-neighbor distance between protons, and $I=1/2$ for $^{1}$H.

In the case of $^{2}$D, which possesses a nuclear quadrupole moment, the relaxation is dominated by the fluctuation of the electric field gradient at the $^{2}$D site and $\expval{\Delta \omega^{2}}$ is described as~\cite{Dimitropoulos1990,Abragam1961}
\begin{equation}
\expval{\Delta \omega^{2}}=\frac{1}{30}\expval{\omega_{Q}^{2}},
\end{equation}
where $\expval{\omega_{Q}}$ denotes the quadrupole coupling frequency given by
\begin{equation}
\omega_{Q}=\frac{3}{2}\frac{e^{2}qQ}{\hbar}.
\end{equation}
Here, $q$ is the electric field gradient.

\begin{figure}[htb]
\includegraphics*[width=8.5cm,trim=0 0 0 0]{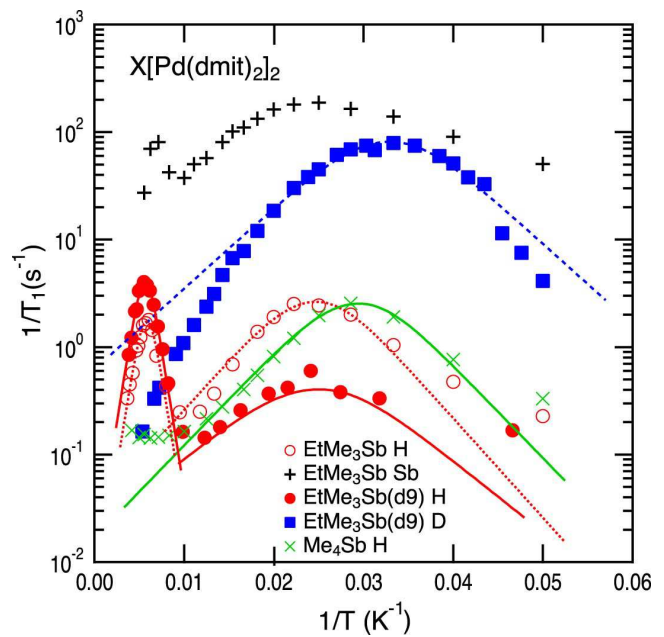}
\caption{Replot of $1/T_{1}$ shown in Fig.~\ref{fig:iT1Tdep}. The solid and dashed lines are fits assuming that $\tau_{c}$ follows the activation temperature dependence $1/\tau_{c}\propto \exp(-\Delta/T)$.}
\label{fig:iT1invT}
\end{figure}

We replot the observed $1/T_{1}$ in Fig.~\ref{fig:iT1invT} as well as fitted curves assuming $\tau_{c}$ that follows the  Arrhenius equation ($1/\tau_{c}\propto \exp(-\Delta/T)$). The enhanced $1/T_{1}$ obtained by $^{2}$D NMR for the d9-deuterated sample peaked at $1/T = 0.033$ K$^{-1}$ ($T = 30$K), and comparing the corresponding suppression of $1/T_{1}$ of $^{1}$H with that of the pristine sample shows that the peak at $T = 30$ K originated from the rotation of the methyl group, whereas that at $T = 200$ K is caused by the rotation of the ethyl group in the cation. 

\begin{figure}[hbt]
\includegraphics*[width=8.5cm]{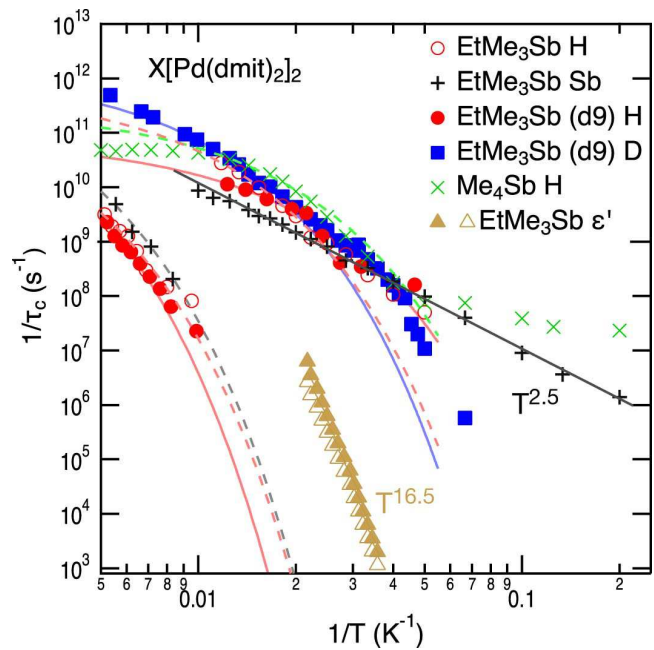}
\caption{Temperature-dependent correlation time $\tau_{c}$ of the cation sites of \textit{X}[Pd(dmit)$_{2}$]$_{2}$ obtained by fitting the data by Eq.~\ref{eq:BPP}. The peak temperature of the dielectric constant as a function of the oscillating frequency (unit of radian) is shown as closed triangles.~\cite{AbdelJawad2013} Corrected microscopic time constants $\tau_{m}$ of the dielectric constants are shown as open triangles. Solid and dashed curves follow the activation temperature dependence. $1/\tau_{c}$ for $^{121}$Sb NMR and the dielectric constant appear linear in the plot and follow an algebraic temperature dependence.}
\label{fig:itauactiv}
\end{figure}

Our main interest is the temperature-dependent low-energy fluctuations of the system and the universal time constants of the electronic correlations at the three nuclear sites. We plot in Fig.~\ref{fig:itauactiv} the correlation time $\tau_{c}$ of $1/T_{1}$ of $^{1}$H, $^{2}$D, and $^{121}$Sb. $1/\tau_{c}$ for $^{1}$H and $^{2}$D almost follow the activation formula
\begin{equation}
1/\tau_{c}=1/\tau_{c0}\exp(-\Delta/T),
\label{eq:activ}
\end{equation}
where $\Delta$ is the activation energy in units of K. The solid and dashed lines in Fig.~\ref{fig:itauactiv} are the fitted plots of $1/\tau_{c}$ to Eq.~\ref{eq:activ}. We summarize the activation energies $\Delta$ (K) in Table~\ref{tab:gap} and find that $\Delta$ predominantly depends on the rotating groups and is insensitive to the nuclei observed.

The rotation of the ethyl group also enhances $1/T_{1}$ of Sb site peaked at $T\sim140$ K as shown in Fig.~\ref{fig:iT1invT}. The obtained $1/\tau_{c}$ above 120 K for $^{121}$Sb well agrees that for $^{1}$H as shown in Fig.~\ref{fig:itauactiv}. This indirect effect to the local fluctuation of the electric field gradient at Sb site indicates that the rotation axis of the ethyl group is a skew line to the Sb-C bonds, by which local field is perturbed. 

Below 100 K, however, $1/\tau_{c}$ for $^{121}$Sb has no characteristic activation energy but slows down gradually, which crosses $1/\tau_{c}$ for $^{2}$D and $^{1}$H originating from the methyl rotation. This shows little indirect coupling between the electric field gradient at the Sb site and the rotation of the methyl group, which is consistent with the $^{2}$D NMR spectra with axial symmetry. The rotation axis agrees to the Sb-CH$_{3}$ bonds and the local rotation is not seriously perturb the electronic state at Sb. This argument is supported by contrasting magnitudes of the enhancements of $1/^{121}T_{1}$ for two peaks, while nearly the same enhancements of $1/^{1}T_{1}$ are observed for the rotations of the ethyl and methyl groups. $1/\tau_{c}$ is fitted by a power relation to the temperature as $\displaystyle 1/\tau_{c} \propto T^{\nu}$ with $\nu=2.5$. This glassy slowing down of $1/\tau_{c}$ well explains the peak of $1/T_{2}$ at 20 K shown in Fig.~\ref{fig:iT1Tdep} because the energy of $1/T_{2}$ is characterized by the dephasing process of the nuclear magnetizations at $\sim 100$ kHz.~\cite{Fujiyama2003b}

\begin{table}
\caption{Activation energies $\Delta$ of the correlation time $\tau_{c}$ of the rotations of methyl and ethyl groups of the pristine and d9-deuterated \textit{X}=EtMe$_{3}$Sb and \textit{X}=Me$_{4}$Sb of \textit{X}[Pd(dmit)$_{2}$]$_{2}$. The rotation axis of the ethyl of d9-deuterated sample does not agree to the Sb-C bonds, which cause a small enhancement of $1/T_{1}$ for $^{1}$H in the ethyl group by indirect coupling. The unit is K. }
\label{tab:gap}
\begin{ruledtabular}
\begin{tabular}{l c| cc|c} 
\multicolumn{1}{c}{} & \multicolumn{1}{c|}{pristine sample} & \multicolumn{2}{c|}{d9-deuterated} & \multicolumn{1}{c}{Me$_{4}$Sb}  \\ 
 & H & H & D & H  \\ \hline
Methyl& 213 & (153) & 172 & 198 \\
Ethyl& 1175 & 1206 & \--- & \---  \\ 
\end{tabular}
\end{ruledtabular}
\end{table}


The nuclear relaxations of the proton and deuteron NMR unveil the rotations of methyl and ethyl groups, whose temperature dependence follows Eq.~\ref{eq:BPP}. For $\beta'$-EtMe$_{3}$Sb[Pd(dmit)$_{2}$]$_{2}$, a similar enhanced dielectric constant was reported in Ref.~\onlinecite{AbdelJawad2013}. Although the enhancement of the dielectric response shows the slow charge dynamics of the sample, it is still under debate whether the magnetically active [Pd(dmit)$_{2}$]$_{2}$ dimers or the cations that is electronically closed cause the slow fluctuation of the electrons.

The dielectric response at the classical limit is expressed as
\begin{equation}
\epsilon(\omega)=  \epsilon_{\infty}+\frac{\epsilon_{0}-\epsilon_{\infty}}{i \omega \tau_{M}}
\label{eq:epsilon}
\end{equation}
Here, $\epsilon_{\infty}$ and $\epsilon_{0}$ denote the dielectric constants at the highest- and lowest-frequency limits, $\tau_M$ is the characteristic time constant for macroscopic polarization, respectively.

Equation~\ref{eq:epsilon} is expected to have a maximum at $\omega \tau_{M} =1$. We plot in Fig.~\ref{fig:itauactiv} the inverted oscillating frequency (unit is rad [s$^{-1}$]) as a function of the peak temperature of the dielectric constant. $1/\tau_{M}$ for the dielectric constant lies between the two values of $1/\tau_{c}$ for the rotations of the ethyl and methyl groups. $1/\tau_{M}$ is $\sim10^{5}$ times smaller than $1/\tau_{c}$ for the rotation of the ethyl group and $\sim 10^{5}$ times larger than $1/\tau_{c}$ of that of the methyl group of the cation. The temperature dependence of $1/\tau_{M}$ follows an algebraic relation with temperature, $1/\tau_{c}\propto T^{\nu}$ with $\nu=16.5$, while the two values of $1/\tau_{c}$ follow the activation temperature dependence. The contrasting values and temperature dependences of $\tau_{M}$ and $\tau_{c}$ show that the electronic fluctuations by the rotations of the methyl and ethyl groups do not directly cause the enhancement of the dielectric constant. We examine the relation between $\tau_{M}$ and microscopic relaxation time $\tau_{m}$ of dipole moments, which should be more favorable for comparison with $\tau_{c}$ measured by NMR. We apply the widely accepted formula $\displaystyle \tau_{M}=3\epsilon_{0}\tau_{m}/(2\epsilon_{0}-\epsilon_{\infty}) $~\cite{Powles1953,Cole1965} and plot $\tau_{m}$ using the values of $\epsilon (\omega)$ at 1 MHz and 316 Hz for $\epsilon_{\infty}$ and $\epsilon_{0}$, respectively. This correction is not ideal, but the correction factor $3\epsilon_{316 \mathrm{Hz}}/(2\epsilon_{316 \mathrm{Hz}}-\epsilon_{1 \mathrm{MHz}})$ shows a moderate change on the order of unity, ranging from 1.7 at 28 K to 2.3 at 46 K. We plot $\tau_{m}$ in Fig.~\ref{fig:itauactiv} as open triangles and find that the correction has a minor effect on the estimated correlation time.

Whereas the motion of the methyl and ethyl groups has little impact on the dielectric response, we can claim a correspondence between the slow charge dynamics and the fluctuation of the local electric field at the Sb site of the EtMe$_{3}$Sb cation. $1/\tau_{c}$ for $^{121}$Sb NMR shows an unconventional slowing down of the local field following the power relation $1/\tau_{c}\propto T^{\nu}$ with $\nu=2.5$. Although the two indices $\nu$ for the dielectric constant and the local fluctuation of the electric field are different, both electric responses exhibit glassy slowing down of multiple electric potentials. This argument again supports the conclusion in Ref.~\onlinecite{AbdelJawad2013} that applied a Fourier-transformed stretched exponential function to analyze $\epsilon'(\omega)$. Shift from the simplest Debye formula shows a glassy electronic state.

The observed algebraic electric fluctuations have the potential to promote the emergence of a QSL. The correlation time of the fluctuation at $^{121}$Sb continues to survive $1/\tau_{c}>10^{6}$ s$^{-1}$, even at $T=5$ K, which can disturb classical magnetic ordering. The power relations of the correlation times to the temperature $1/\tau_{c,M}\propto T^{\nu}$ observed by NMR and the dielectric constants have no finite critical temperature, suggesting the existence of the quantum critical point at zero temperature. In addition, it has been theoretically well established that quantum liquids in one and two dimensions are characterized by algebraic electronic correlations, and our results are in line with this understanding.~\cite{Po2015,Haldane1981} While the origin of the observed glassy state is still open, we comment that nearly degenerate energy levels in a QSL are on the verge of algebraic particle excitations and a concomitant glassy state in space and time domains.

\section{Summary}

In this report, we demonstrated the nuclear spin-lattice and spin-spin relaxations of the cation of the candidate quantum spin-liquid $\beta'$-EtMe$_{3}$Sb[Pd(dmit)$_{2}$]$_{2}$. Significant enhancements of $1/T_{1}$ and $1/T_{2}$ were observed at $^{1}$H and $^{121}$Sb in the pristine sample and at $^{1}$H, $^{2}$D in the d9-deuterated samples. The correlation times follow the activation temperature dependence, and we estimate the activation energies $\Delta=2\times10^{2}$ K ($1.2\times 10^{3}$ K) for the rotation of the methyl (ethyl) group respectively.

An unconventional charge dynamics was found for the Sb site in the cation. The algebraic slowing down of the charge dynamics with a small exponent, $1/\tau_{c}\propto T^{2.5}$, shows that the charge fluctuation remains active even at the lowest temperature, which can prevent classical magnetic ordering and stabilize QSL. The observation of the algebraic charge dynamics shows that charge correlation as well as magnetic correlation can be strongly modified in quantum liquids.

\begin{acknowledgments}
We are grateful to K.~Ueda for helpful discussions. This work was supported by Grants-in-Aid for Scientific Research (C) (26400378) and (S) (16H06346) from JSPS.
\end{acknowledgments}
\end{document}